\documentclass[prd,showpacs,amsmath,amssymb,twocolumn]{revtex4-1}
\usepackage{amssymb,bm}
\usepackage{amsfonts}
\usepackage[misc]{ifsym}
\usepackage{authoraftertitle}
\usepackage{overpic,graphicx}
\usepackage{keyval,graphicx}
\usepackage{textcomp,wasysym}
\usepackage[perpage,symbol]{footmisc}%
\usepackage[CJKbookmarks=true, colorlinks=true,
linkcolor=blue, urlcolor=blue,citecolor=blue]{hyperref}
\usepackage[english]{babel}
\begin{document}

\title{Evaluation of the $\theta_K$ and the Mixing angle $\theta _{h_1}$}
\author{Ya Gao$^{1}$, Yateng Zhang$^{2}$,
Bo Zheng$^{1*}$, Zhen-Hua Zhang$^{1*}$, Wenbiao Yan$^{2}$, Xiaohua Li$^{1}$}

\affiliation{1, School of Nuclear Science and Technology, University of South China, Hengyang, Hunan 421001, China}
\affiliation{2, University of Science and Technology of China, Hefei, Anhui 230026, China}

\begin{abstract}
	The $\theta_K$ has been re-evaluated via mass relations
	and latest experimental results, meanwhile, the $M_{K_{1B}}$ also be obtained.
	Based on the singlet-octet mixing model and quark-flavor mixing model,
	the $\theta_{h_1}$ has been recalculated with a modified formula, by inputting $M_{K_{1B}}$
	instead of $\theta_K$. The values are calculated to be
	$|{\theta _K}| = {(42.6 \pm 2.2)^ \circ }$, $M_{K_{1B}} = 1333.9 \pm 4.6\ {\rm{MeV}}/c^{2}$
	and $\theta _{h_1} = 29.5\pm 2.0^\circ$. Using these calculations as input, we predict
	that the branching fraction of $J/\psi  \to \eta'(\eta)h_1(1170)$ is about
	one order higher than that of $J/\psi  \to \eta'(\eta)h_1(1415)$, which can be measured
	in future experiments to test the validity of these two models.
\end{abstract}

\maketitle
\setlength{\parskip}{0.5\baselineskip}
\section{Introduction}\vspace*{-2mm}
\label{sec:intro}
In the quark model, pure flavor singlet and octet can be mixed to form mass
eigenstates due to flavor SU(3) symmetry breaking.
A parameter, octet-singlet mixing angle $\rm{\theta}$, is used to construct
the mixing formation.
Since mixing angle is tightly related with decoupling of $u$ and
$d$ quarks from $s$ quark, the determination of mixing angle in
different types of singlet and octet, especially for vector and tensor mesons,
is extremely significant for us to understand hadronic physics.

There are two kinds of P-wave axial-vector mesons with $J^{P}=1^{+}$,
which are $\rm{^3P_1}$ and $\rm{^1P_1}$, respectively.
According to experimental results, $a_1(1260)$, $f_1(1285)$ and
$f_1(1420)$ are members of $\rm{^3P_1}$ state with $J^{PC}=1^{++}$,
while $b_1(1235)$, $h_1(1170)$ and $h_1(1415)$, which called as $h_1(1380)$
previously and re-named as $h_1(1415)$ by PDG~\cite{pdg2019}, are members of $\rm{^1P_1}$
state with $J^{PC}=1^{+-}$.
Besides, $K_{1A}$ and $K_{1B}$ are $\rm{^3P_1}$ and $\rm{^1P_1}$ states, respectively, with $J^{P}=1^{+}$ without definite $C$ parity.
Among them, $f_1(1285)$ was firstly discovered in 1965 by CERN~\cite{ref1} and BNL~\cite{ref2}.
Since then, many experimental investigations have been performed on $f_1(1285)$
and later discovered $f_1(1420)$.
Their masses were predicted to be $M_{f_1(1285)}=1229.5\ {\rm{MeV}}/c^{2}$ and
$M_{f_1(1420)}=1487\ {\rm{MeV}}/c^{2}$~\cite{ref3} based on hypothesis of quark flavor states.
However, these predictions are inconsistent with the averaged experimental measured values, with more than $5\sigma$.
The same thing happened to $h_1(1170)$ and $h_1(1415)$.
To incorporate these states into quark model, a mixing method are employed to explain them.
There are two mixing schemes to solve this problem, which is singlet-octet (SO) basis and
quark-flavour (QF) basis~\cite{ref4,ref5,ref6,ref7,ref8}, respectively.
Actually, these two mixing models are essentially the same since both of them are constructed
on the basis of mass differences between $u$, $d$ and $s$ quarks.
By the way, there are also other explanations on the $h_1(1170)$ and $h_1(1415)$, e.g.,
pseudoscalar-vector meson interaction~\cite{lwh1,lwh2}.

In the P-wave state axial-vector meson members, $f_1(1285)$, $f_1(1420)$, $h_1(1170)$ and $h_1(1415)$ are
regarded as partial strange axial-vector mesons, $K_{1A}$ and $K_{1B}$ mesons are strange partners of
$a_1(1260)$ and $b_1(1235)$, respectively, while $a_1(1260)$ and $b_1(1235)$ are neutral~\cite{ref9}.
The mixing between states of $\rm{^3P_1}$ and $\rm{^1P_1}$ with definite $C$-parity are forbidden,
which makes the mixing between $K_{1A}$ and $K_{1B}$ to be the unique place to study the relationship between them.
The mixing angle between $K_{1A}$ and $K_{1B}$, named as $\theta_K$, is responsible for
the physical mass eigenstates $K_1(1270)$ and $K_1(1400)$. Therefore the
precise determination of $\theta_K$ will directly contribute to the estimation of
$K_{1A}$ and $K_{1B}$ masses, and it is also of great importance to the study of P-wave
mesons. Furthermore, its bridge function on understanding the mass relation of the
vector mesons flavor-SU(3) singlet and octet states due to mixing, is of fundamental importance on hadronic physics.
Besides, $\theta_K$ is an input parameter when studying production rates of $K_1(1270)$ and $K_1(1400)$
in $B$ and $D$ decays~\cite{ref10,ref11,ref12,ref13,ref14,ref15,ref16,ref17,ref18,ref19}, which also calls for
more careful studies.

This paper is mainly divided into four parts. Firstly, we present experimental results
on P-wave states axial-vector mesons. Secondly, we discuss the $\theta_K$ and $K_{1B}$ mass via a mass-squared matrix.
Thirdly, we used the $h_1(1415)$ masses measured by LASS~\cite{ref20}, Crystal Barrel~\cite{ref21} and
BESIII~\cite{ref22,ref23} to calculate the mixing angle of $h_1(1170)-h_1(1415)$.
In the last section, we use calculated mixing angle to predict production ratio of
${J/\psi\to\eta'h_1(1415)}$ and ${J/\psi\to\eta'h_1{(1170)}}$.

\section{Experimental results}
\label{sec:expr}
Currently, the masses of $f_1(1285)$ and $f_1(1420)$ are measured via various decay model,
such as $\eta \pi \pi$, $\gamma {\rho ^0}$~\cite{ref24}, four-pion~\cite{ref25,ref26},
and $K\bar K\pi$~\cite{ref27}. Their averaged values are $M_{f_1(1285)} = 1281.9\pm0.5\ {\rm{MeV}}/c^{2}$
and $M_{f_1(1420)} = 1426.3 \pm 0.9\ {\rm{MeV}}/c^{2}$~\cite{ref9}, respectively.
The mixing angle between $f_1(1285)-f_1(1420)$ is regarded to be about $50^\circ$ based
on analysis of generalized Gell-Mann-Okubo mass relation~\cite{ref28}, SU(3) coupling
formula, $\gamma\gamma^{*}$ decays of $f_1(1285)$ and $f_1(1420)$~\cite{ref29}, and radiative
${J/\psi}$ decays studied by Close and Kirk~\cite{ref30}.

The latest study about $h_1(1170)$ was performed by A.~Ando {\it et al.}~\cite{ref31}.
They carried out a partial wave analysis for the system of ${\pi ^ + }{\pi ^ - }{\pi ^0}$,
and the mass of $h_1(1170)$ is measured to be $M = 1168 \pm 4\ {\rm{MeV}}/c^{2}$, which is consistent
with the estimated value $M = 1170 \pm 20\ {\rm{MeV}}/c^{2}$ from PDG~\cite{ref9}.
Theoretically, the meson mixing model~\cite{ref29,ref32} predicts the
mass of $h_1(1415)$ to be $M_{h_1(1415)} = (1489.75\pm18.08)\ {\rm{MeV}}/c^2$ or $M_{h_1(1415)} = 1468\ {\rm{MeV}}/c^2$,
while predictions of models which based on quark antiquark and chiral symmetry breaking effects lie in a wide range,
from $1386.42\ {\rm{MeV}}/c^2$ to $1511\ {\rm{MeV}}/c^2$~\cite{ref33,ref34,ref35,ref36,ref37}.
On the experimental side, LASS~\cite{ref20},
Crystal Barrel~\cite{ref21} and BESIII~\cite{ref22,ref23} have measured the mass of $h_1(1415)$ with
large uncertainty, which are summarized in Tab.~\ref{tab:weight}.
In the table, significant deviations and large uncertainty on different experimental measured
$h_1(1415)$ masses before 2018 are shown, which leads to a controversy on its nature and no
assignment on this particle. Benefit from the precisely measured mass by
BESIII at 2018, the PDG began to record it as a normal meson and assigned as $h_1$ state.

\begin{table}[htbp]
	\caption{\small Results of $h_1(1415)$ mass and $\theta_{h_1}$.}
	\begin{center}
		\begin{tabular}{l|cc }\hline
			Experiments   &${M_{h_1(1415)}}({\rm{MeV/}}{c^{2}})$ & ${\theta_{h_1}}$        \\ \hline
			LASS          &$1380\pm20$&$(18.4\pm7.2)^{\circ}$     \\
			Crystal Barrel&$1440\pm60$&$(35.8\pm14.4)^{\circ}$  \\
			BESIII2015    &$1412.2\pm8.9$ &$(26.8\pm4.2)^{\circ}$   \\
			BESIII2018    &$1423.2\pm7.6$&$(29.5\pm2.0)^{\circ}$  \\
			Average       &$1418.1\pm7.1$&$(28.3\pm3.7)^{\circ}$   \\
			\hline
		\end{tabular}
		\label{tab:weight}
	\end{center}
\end{table}

\section{Discussion on $\pmb{\theta_K}$}
\label{sec:thetaK}

According to Appelquist-Carazzone decoupling theorem \cite{ref38}, one can
infer that when the $s$ quark mass ${m_s}$ becomes as large as comparable with $\rm{\Lambda_{QCD}}$,
which represents the scale of confinement and spontaneous chiral symmetry breaking in quantum chromodynamics (QCD), it is
possible to form an effective low-energy theory with $s$ quark decoupled completely.
Therefore, two mass eigenstates, one composed of the light $u$ and $d$
quarks and orthogonal to the which other only composed of $s$ quark, must be produced
by the mixing of meson flavor eigenstates.
In the case of $f_1(1285)$ and $f_1(1420)$ or $h_1(1170)$ and $h_1(1415)$,
from the mass relation, the mixing is accomplished through a mixing angle which depends on the masses of
$K_{1A}$ and $K_{1B}$ which in turn depends on $\theta_K$.
Besides, under the $\rm{SU(3)}$ gauge theory \cite{ref28}, the decoupling angle $\theta_{0}=35.3^{\circ}$ is
considered as an ideal mixing angle for vector or tensor mesons.

The first study on $\theta_K$ was performed by Carnegie {\it et al} in 1977~\cite{ref39}, and
the value is calculated to be $(41\pm 4)^\circ$. Thereafter, many phenomenological analyses
explore the value of $\theta_K$ via the processes involving
$K_1$ mesons. H. Hatanaka and K. C. Yang suggested that the $\theta_K$ favors $(34 \pm 13)^\circ$
through considering the branching fraction of $B \to  K_1\gamma$ as the function of
$\theta_K$~\cite{ref10}. Under the framework of QF basis, M. Suzuki predicted that
$\theta_K\simeq33^\circ $ or $ 57^\circ$ through the study of $\tau\to K_1\mu$~\cite{ref40} while
F. Divotgey group gave $\rm{|\theta_K|=(33.6 \pm 4.3)^\circ}$ based on the study of
$K_1\to K^{*}\pi/K\rho/K\omega$~\cite{ref41}, respectively. Under the SO basis,
as well as essentially identical QF basis, two different results are expected to be
$\theta_K\sim34^\circ$ by H.~Y.~Cheng and $|\theta_K|\simeq(59.55\pm2.81)^\circ$ by
D.~M.~Li {\it et al} in two kinds of mass relations~\cite{ref29,ref42}, respectively.
In addition to these, the Relativized Quark Model~\cite{ref34,ref43} and
Nonrelativistic Constituent Quark Model(NCQM)~\cite{ref35,ref44,ref45} are employed to
predict $\theta_K$ and the resultant values are in the range from $31^\circ$ to $55^\circ$.
Besides, some other models are proposed to study
$\theta_K$, such as Quark-Pair-Creation Model(QPCM)~\cite{ref46} and Dynamical Model~\cite{ref47},
and the predicted results are $\theta_K=45^\circ$ and about $60^\circ$, respectively.
The $\theta_K$ values calculated by various models are summarized in Table \ref{tab:thetaK}.

Under QF basis and SO basis, the mixing between $f_1(1285)$ and $f_1(1420)$
or $h_1(1170)$ and $h_1(1415)$ can be written as~\cite{ref41}:

\begin{widetext}
	\begin{equation}
	\left( {\begin{array}{*{20}{c}}
		{{f_1}(1285)}\\
		{{f_1}(1420)}
		\end{array}} \right) = \left( {\begin{array}{*{20}{c}}
		{\cos {\theta _{+}}}&{\sin {\theta _{+}}}\\
		{ - \sin {\theta _{+}}}&{\cos {\theta _{+}}}
		\end{array}} \right)\left( {\begin{array}{*{20}{c}}
		1\\
		8
		\end{array}} \right) = \left( {\begin{array}{*{20}{c}}
		{\cos {\alpha _{+}}}&{\sin {\alpha _{+}}}\\
		{ - \sin {\alpha _{+}}}&{\cos {\alpha _{+}}}
		\end{array}} \right)\left( {\begin{array}{*{20}{c}}
		N\\
		S
		\end{array}} \right)\
	\end{equation}
	
	and
	
	\begin{equation}
	\left( {\begin{array}{*{20}{c}}
		{h_1}(1170)\\
		{h_1}(1415)
		\end{array}} \right) = \left({\begin{array}{*{20}{c}}
		{\cos {\theta _{-}}}&{\sin {\theta _{-}}}\\
		{ - \sin {\theta _{-}}}&{\cos {\theta _{-}}}
		\end{array}} \right)\left( {\begin{array}{*{20}{c}}
		{{1}}\\
		{{8}}
		\end{array}} \right) = \left( {\begin{array}{*{20}{c}}
		{\cos {\alpha _{-}}}&{\sin {\alpha _{-}}}\\
		{ - \sin {\alpha _{-}}}&{\cos {\alpha _{-}}}
		\end{array}} \right)\left( {\begin{array}{*{20}{c}}
		N\\
		S
		\end{array}} \right),\
	\end{equation}
\end{widetext}
where ${1 = (u\bar u + d\bar d + s\bar s)/\sqrt 3}$,
${8 = (u\bar u + d\bar d - 2s\bar s)/\sqrt 6}$,
${N = (u\bar u + d\bar d)/\sqrt 2}$ and ${S = s\bar s}$, ${\theta}$ and ${\alpha}$ are mixing angle under the SO basis and QF basis,
subscript $+/-$ represents the $\rm{^3P_1}$/$\rm{^1P_1}$ states, respectively.
The relationship of between ${\alpha}$ and ${\theta}$ is
${\alpha  = \theta_{0} - \theta}$.

Generally, the mixing between $\rm{^3P_1}$ and $\rm{^1P_1}$ states can be expressed as
\begin{equation}
{M_{{K_{1A}}}^2 = M_{{K_1}(1400)}^2{\cos ^2}{\theta _K} + M_{{K_1}(1270)}^2{\sin ^2}{\theta _K}},
\end{equation}
\begin{equation}
{M_{{K_{1B}}}^2 = M_{{K_1}(1400)}^2{\sin ^2}{\theta _K} + M_{{K_1}(1270)}^2{\cos ^2}{\theta _K}}.
\end{equation}
From mass relations~\cite{ref29,ref42}, the mixing angle ${\theta _+}$ and
${\theta _-}$ rely on $K_{1A}$ and $K_{1B}$ masses, respectively. From equation (3) and (4),
it is clear that the mass of $K_{1A}$ or $K_{1B}$ is dependent on the mixing angle between $K_{1A}$ and $K_{1B}$, ${\theta _K}$.

Since no great improved measurement on $h_1(1415)$ in the foreseeable future
than that by BESIII, we update this result by inputting
the BESIII measured $M_{{h_1}(1415)} = 1423.2 \pm 7.6\ {\rm{MeV}}/c^{2}$ to calculate $\theta_K$ and $M_{K_{1B}}$,
and the results are
\begin{eqnarray}
&|\theta _K| = (42.6 \pm 2.2)^ \circ ,&\\
&M_{K_{1B}} = 1333.9 \pm 4.6\ {\rm{MeV}}/c^{2}.&
\end{eqnarray}
The $\theta_K$ is consistent with most theoretical predictions which hold the
$\theta_K$ is smaller than $45^{\circ}$.

\begin{table}
	\caption{\small Theoretical predictions of $\theta_K$.}
	\begin{center}
		\begin{tabular}{l|cc}\hline
			
			Models         &Results \\ \hline
			Carnegie et al.~\cite{ref39}&  ${\theta_K=(41\pm4)^\circ}$\\
			H.Hatanaka et al.~\cite{ref10}&  ${\theta_K=(34\pm13)^\circ}$\\
			QF basis~\cite{ref40,ref41}   & ${\theta_K\sim33^\circ/57^\circ}$\\
			& ${|\theta_K|=(33.6 \pm 4.3)^\circ}$\\
			SO basis~\cite{ref29,ref42} &${\theta_K\sim34^\circ}$\\
			& ${|\theta_K|=(59.55 \pm 2.81)^\circ}$\\
			Relativized Quark Model~\cite{ref34,ref43}& ${\theta_K\sim34^\circ}$\\
			&${\theta_K\simeq45^\circ}$\\
			NCQM~\cite{ref35,ref44,ref45} & ${35^\circ\le\theta_K\le55^\circ}$\\
			&${\theta_K=(31\pm4)^\circ}$ \\
			&${\theta_K=(37.3\pm3.2)^\circ}$\\
			
			QPCM~\cite{ref46} &${\theta_K\simeq60^\circ}$\\
			Dynamical Model~\cite{ref47} &${\theta_K=45^\circ}$\\\hline
		\end{tabular}
		\label{tab:thetaK}
	\end{center}
\end{table}

\section{Discussion on the $\pmb{h_1(1415)-h_1(1170)}$ Mixing angle}
\label{sec:hmix}

Given the most precise $h_1(1415)$ mass measured by BESIII, we update the calculation
on $\theta_{h_1}$, by only rely on $M_{K_{1B}}$ obtained in section~\ref{sec:thetaK}, which differs
to previous models with additional $\theta_K$ uninvolved.

In this section, we use experimental results of $h_1(1415)$ mass from LASS~\cite{ref20}, Crystal
Barrel~\cite{ref21} and BESIII collaboration~\cite{ref22,ref23} to calculate the $\theta_{h_{1}}$,
and an averaged $\theta_{h_{1}}$ is also presented.

Under the SO model, the mixing of $h_1(1170)$ and $h_1(1415)$ can be described by a symmetric matrix as~\cite{ref41}
\begin{widetext}
	\begin{equation}
	{{M^2}} = \left( {\begin{array}{*{20}{c}}
		{{m_1^2}}&{\delta} \\
		\delta &{{m_8^2}}
		\end{array}} \right) = \left( {\begin{array}{*{20}{c}}
		{{m_L^2 + m_H^2 - m_8^2}}&{ - {{\left( {{m_8^2(m_L^2 + m_H^2 - m_8^2) - m_L^2m_H^2}} \right)}^{1/2}}}\\
		{ - {{\left( {{m_8^2(m_L^2 + m_H^2 - m_8^2) - m_L^2m_H^2}} \right)}^{1/2}}}&{{m_8^2}}
		\end{array}} \right),
	\end{equation}
\end{widetext}
where $m_1$ and $m_8$ represent the masses of singlet and octet, respectively.

According to Ref.~\cite{ref41}, one have
\begin{equation}
{\cos 2\theta}  = \frac{{{m_8^2 - m_1^2}}}{{{m_H^2 - m_L^2}}},~
{\tan \theta}  = \frac{{{m_8^2 - m_H^2}}}{\delta }.
\end{equation}
Basing on (7) and (8), the relation between mixing angle and the masses can be written as
\begin{equation}
{\tan\theta}=\frac{{m_H^{2}-m_8^{2}}}{\sqrt{{m_8^{2}(m_H^{2}+m_L^{2}-m_8^{2})-m_H^{2}m_L^{2}}}},
\end{equation}
where the mass of the octet state can be deduced from Gell-Mann-Okubo relations~\cite{ref28} as
\begin{equation}	
{m_8^2} = \frac{1}{3}({4m_{{K_{1B}}}^2 - m_{{b_1}(1235)}^2}).
\end{equation}
With some appropriate replacement such as ${{m_H}} = {{M_{{h_1}(1415)}}}$ and
${{m_L}} = {{M_{{h_1}(1170)}}}$, and inputting the result of ${{M_{h_1(1415)}}}$
from three experiments mentioned above, the ${\theta_{h_1}}$ can be calculated, and the
results are shown in Tab.~\ref{tab:weight}. The averaged ${\theta_{h_1}}$
is also shown in Tab.~\ref{tab:weight}.

\section{Branching fraction about $\pmb{{J/\psi\to\eta'h_1(1415)/h_1(1170)}}$}

The re-calculated values of $\theta_{h_{1}}$ give us an opportunity to estimate the production rates of channels involving $h_1(1415)$ and $h_1(1170)$.

The decay widths of $J/\psi  \to {\eta'}{h_1}$ and $J/\psi  \to {\eta}{h_1}$ can be written as
\begin{equation}
{{\Gamma _{J/\psi  \to {\eta'}{h_1}}}} = \frac{{{|{q_{{h_1}}}{|^3}}}}{{{8\pi m_{{\eta'}}^2}}}|\frac{{{\mathcal{A}_{J/\psi  \to {\eta'}{h_1}}}}}{{{\varepsilon} \cdot{q_{J/\psi }}}}{|^2}
\end{equation}
and
\begin{equation}
{{\Gamma _{J/\psi  \to {\eta}{h_1}}}} = \frac{{{|{q_{{h_1}}}{|^3}}}}{{{8\pi m_{{\eta}}^2}}}|\frac{{{\mathcal{A}_{J/\psi  \to {\eta'}{h_1}}}}}{{{\varepsilon} \cdot{q_{J/\psi }}}}{|^2},
\end{equation}

where ${h_1}$ presents ${h_1}(1170)$ or ${{h_1}(1415)}$, ${m_{{\eta'}}}$ and ${m_{{\eta}}}$ are masses of
${\eta'}$ and ${\eta}$, respectively, ${q}$ is momentum and ${\varepsilon}$ is the polarization vector of $J/\psi$.
${\mathcal{A}}$ is the amplitude of the process and can be expressed as
\begin{equation}
{\mathcal{A} =  < {h_1(1170)}{\eta'}|\hat H|J/\psi  >},
\end{equation}

\begin{equation}
{\mathcal{A} =  < {h_1(1415)}{\eta'}|\hat H|J/\psi  >}
\end{equation}
in $J/\psi  \to {\eta'}{h_1}$ and

\begin{equation}
{\mathcal{A} =  < {h_1(1170)}{\eta}|\hat H|J/\psi  >},
\end{equation}

\begin{equation}
{\mathcal{A} =  < {h_1(1415)}{\eta}|\hat H|J/\psi  >}
\end{equation}

in $J/\psi  \to {\eta}{h_1}$, respectively.

Whether in process $J/\psi  \to {\eta'}{h_1}$ or $J/\psi  \to {\eta}{h_1}$, under QF basis and SO basis, the production amplitudes of decay ${h_1}(1415)$ and ${{h_1}(1170)}$ can be expressed as
\begin{equation}
{\mathcal{A}({h_1}(1170)) = (\sqrt 2 \cos \alpha  + R\sin \alpha ){g_0}},
\end{equation}
\begin{equation}
{\mathcal{A}({h_1}(1415)) =( - \sqrt 2 \sin \alpha  + R\cos \alpha ){g_0}},
\end{equation}
and
\begin{equation}
{\mathcal{A}({h_1}(1170)) = \left[ {\frac{{(2 + R)\cos \theta }}{{\sqrt 3 }} + \frac{{(2 - 2R)\sin \theta }}{{\sqrt 6 }}} \right]{g_0}},
\end{equation}
\begin{equation}
{\mathcal{A}({h_1}(1415)) = \left[ {\frac{{(2 - 2R)\cos \theta }}{{\sqrt 6 }} - \frac{{(2 + R)\sin \theta }}{{\sqrt 3 }}} \right]{g_0}},
\end{equation}
respectively, where ${\theta}$ and $\alpha$ are set to be ${29.5^ \circ }$ and $5.8^\circ$,respectively, according to the prediction of section~\ref{sec:hmix},
means mixing angle of
${h_1(1170)-h_1(1415)}$ in QF basis, which is ${6^ \circ }$ according to the
section~\ref{sec:thetaK}, ${g_0}$ is coupling constant describing the decay strength,
and $R$ presents the SU(3)-breaking ratio with the value of $0.6298 \pm 0.00068$ \cite{ref29}.
After cancelling the common parameters, the ratio of $Br(J/\psi  \to {\eta'}{h_1}(1415))$ and
$Br(J/\psi  \to {\eta'}{h_1}(1170))$ is only related to momentum and amplitude of $h_1(1415)$
and $h_1(1170)$, and the ratio is
\begin{widetext}
	\begin{equation}
	{\frac{Br(J/\psi\to\eta'(\eta)h_1(1170))}{Br(J/\psi\to\eta'(\eta)h_1(1415))}=
		({\frac{q_{h_1(1170)}}{q_{h_1(1415)}}})^{3}(\frac{\sqrt{2}\cos
			\alpha+R\sin\alpha}{-\sqrt{2}\sin\alpha+R\cos\alpha})^{2}}
	\end{equation}
	in QF basis and
	\begin{equation}
	{\frac{Br(J/\psi\to\eta'(\eta)h_1(1170))}{Br(J/\psi\to\eta'(\eta)h_1(1415))}=
		({\frac{q_{h_1(1170)}}{q_{h_1(1415)}}})^{3}(\frac{\sqrt{6}(2+R)\cos\theta+
			\sqrt{3}(2-2R)\sin\theta}{-\sqrt{6}(2+R)\sin\theta+\sqrt{3}(2-2R)\cos\theta})^{2}}
	\end{equation}
\end{widetext}
in SO basis, respectively. Where $q_{{h_1}(1415)}$ and $q_{{h_1}(1170)}$ are
momenta of $h_1(1415)$ and $h_1(1170)$ in the ${J/\psi}$ rest frame, and written as~\cite{ref9}
\begin{equation}
{q = \frac{[(M^{2}-(m_1+m_2)^{2})(M^{2}-(m_1-m_2)^{2})]^{1/2}}{2M}},
\end{equation}
where $M$ is mass of $J/\psi$, $m_1$ and $m_2$ represent the masses of $\eta^{,}$ or $\eta$ and $h_1(1170)$ or $h_1(1415)$, respectively.
Inputting mixing angles of $h_1(1415)-h_1(1170)$,
in both models the ratio is calculated to be
\begin{equation}
{\frac{Br(J/\psi\to\eta'h_1(1170))}{Br(J/\psi\to\eta'h_1(1415))}\simeq 12}
\end{equation}
and
\begin{equation}
{\frac{Br(J/\psi\to\eta h_1(1170))}{Br(J/\psi\to\eta h_1(1415))}\simeq 13}.
\end{equation}

\section{Summary}

In summary, we have re-evaluated the $\theta_K$, and the calculated value is $\theta _K = (42.6 \pm 2.2)^ \circ $,
which is consistent with the popular opinion that it should be less than $45^\circ$.
The $\theta_{h_1}$ also has been recalculated with a modified formula and a precise result,
$\theta _{h_1} = 29.5 \pm 2.0^\circ$, is obtained. Using this result, the production
of $J/\psi\to\eta'(\eta)h_1(1170)$ is estimated to be about one order higher than
$J/\psi\to\eta'(\eta)h_1(1415)$, which can be tested in the future experiments.

\section{Acknowledgments}

The authors would like to thanks the useful discussion with Prof. Demin Li and Dr. Qifang Lv.
This work was partially supported by National Natural Science Foundation of China
(Project Nos. U1732263, 11975118, 11575077 and 11705081), National Natural Science Foundation
of Hunan Province (Project No. 2019JJ30019).


\begin{thebibliography}{99}
	\bibliographystyle{unsrt}
	\bibitem{pdg2019} M. Tanabashi et al. (Particle Data Group), "2019 Review of Particle Physics," $Physical\ Review\ D,$ vol. 98, no. 03, p.030001, 2018 and 2019 update.
	\bibitem{ref1} C. d'Andlau, J. Barlow, and A. M. Adamson, "Evidence for a non-strange meson of mass 1290 MeV," $Physics\ Letters,$ vol. 17, no. 3, p. 347, 1965.
	\bibitem{ref2} D. H. Miller et al., "$KK\pi$ Resonance at 1280 MeV," $Physical\ Review\ Letters,$ vol. 14, no. 26, p. 1074, 1965.
	\bibitem{ref3} H. Yu, Q. X. Shen, "Properties of the tensor mesons $f_2(1270)$ and $f'_2(1525)$," $Journal\ of\ Physics\ G,$ vol. 27, no.4, p.807, 2001.
	\bibitem{ref4} G. Gidal et al. "Observation of Spin 1 $f_1 (1285)$ in the reaction $\gamma\gamma^{*}\to \eta^0\pi^+\pi^-$," $Physical\ Review\ letters,$ vol. 59, no. 18, p. 2012, 1987.
	\bibitem{ref5} W. S. Carvalho, A. S. de Castro, and A. C. B. Antunes, "SU(3) mixing for excited mesons," $Journal\ of\ Physics\ A,$ vol. 35, no. 35, p. 7585, 2002.
	\bibitem{ref6} K. C. Yang, "Form factors of B(u, d, s) decays into p-wave axial-vector mesons in the light-cone sum rule approach," $Physical\ Review\ D,$ vol. 78, no. 3, p.034018, 2008.
	\bibitem{ref7} J. J. Dudek et al., "Isoscalar meson spectroscopy from lattice QCD, "$Physical\ Review\ D,$ vol. 83, no. 11, p.111502, 2011.
	\bibitem{ref8} J. J. Dudek et al., "Toward the excited isoscalar meson spectrum from lattice QCD," $Physical\ Review\ D,$ vol. 88, no. 9, p. 094505, 2013.
	\bibitem{lwh1} W. H. Liang, S. Sakai, E. Oset, "Theoretical description of the $J/\psi \to \eta(\eta') h_1(1380), J/\psi \to \eta(\eta')h_1(1170)$ and $J/\psi \to \pi^0 b_1(1235)^0$ reactions," $Physical\ Review\ D,$ vol. 99, no. 09, p.094020, 2019.
	\bibitem{lwh2} S. J. Jiang, S. Sakai, W. H. Liang, E. Oset, "The $\chi_{cJ}$ decay to $\phi K^* \bar{K}, \phi h_1(1380)$ testing the nature of axial vector meson resonances," $Physics\ Letters\ B,$ vol. 797, p. 134831, 2019.
	\bibitem{ref9} Tanabashi, Masaharu, et al. "Review of particle physics," $Physical\ Review\ D,$ vol. 98, no. 3, p. 030001, 2018.
	\bibitem{ref10} H. Hatanaka and K. C. Yang, "$B\to K_1\gamma$ decays in the light-cone QCD sum rules,"  $Physical\ Review\ D,$ vol. 77, no. 9, p. 094023, 2008.
	\bibitem{ref11} E. Kou, A. Le Yaouanc and A. Tayduganov, "Determining the photon polarization of the $b\to s \gamma$ using the $B\to K_1(1270)\gamma\to(K\pi\pi)\gamma$ decay," $Physical\ Review\ D,$ vol. 83, no. 9, p. 094007, 2011.
	\bibitem{ref12} M. Sayahi and H. Mehraban, "$B\to J/\psi K(1270)$ and $B\to J/\psi K(1400)$ decays in QCD factorization approach," $Adv\ High\ Energy\ Physics,$ vol. 2012, p. 494031, 2012.
	\bibitem{ref13} X. Liu and Z. J. Xiao, "Branching ratios of $B_c\to A P$ decays in the perturbative QCD approach," $Physical\ Review\ D,$ vol. 81, no. 7, p. 074017, 2010.
	\bibitem{ref14} X. Liu and Z. J. Xiao, "Studies on charmless hadronic $B_c\to VA, AA$ decays in the perturbative QCD approach," $Journal\ of\ Physics\ G,$ vol. 38, no. 3, p. 035009, 2011.
	\bibitem{ref15} H. Y. Cheng, C. K. Chua and C. W. Hwang, "Covariant light-front approach for s-wave and p-wave mesons: Its application to decay constants and form factors," $Physical\ Review\ D,$ vol. 69, no. 7, p. 074025, 2004.
	\bibitem{ref16} J. P. Lee, "Radiative $B\to K_1$ decays in the light-cone sum rules," $Physical\ Review\ D,$ vol. 74, no. 7, p. 074001, 2006.
	\bibitem{ref17} C. H. Chen et al., "Production of $K_0^*(1430)$ and $K_1$ in B decays," $Physical\ Review\ D,$ vol. 72, no. 5, p. 054011, 2005.
	\bibitem{ref18} A. N. Kamal and R. C. Verma, Phys. Rev. D 45, 982 (1992).
	\bibitem{ref19} H. Y. Cheng and C. W. Chiang, "Hadronic D decays involving even-parity light mesons," $Physical\ Review\ D,$ vol. 81, no.7, p. 074031, 2010.
	\bibitem{ref20} D. Aston et al. (LASS Collaboration), "Evidence for two Strangeonium resonances with $J^{PC} = 1^{++}$ and $1^{+-}$ in $K^{-}p$ interactiona at 11 GeV/$c$," $Physics\ letters\ B,$ vol. 201, no.4 , p. 573-578, 1988.
	\bibitem{ref21} A. Abele et al. (Crystal Barrel Collaboration), "Antiproton-proton annihilation at rest into $K_LK_S\pi^0\pi^0$," $Physics\ Letters\ B,$ vol. 415, no. 3, p. 280-288, 1997.
	\bibitem{ref22} M. Ablikim et al. (BESIII Collaboration), "Study of  $\chi_{cJ}$ decaying into $\phi K^{*}(892)K^{-}$," $Physical\ Review\ D,$ vol. 91, no. 11, p. 112008, 2015.
	\bibitem{ref23} M. Ablikim et al. (BESIII Collaboration), "Observation of $h_1(1380)$ in the $J/\psi\to\eta^{,}KK^-\pi$ decay," $Physical\ Review\ D,$ vol. 98, no. 7, p. 072005, 2018.
	\bibitem{ref24} D. Barberis et al. (WA102 Collaboration), "A measurement of the branching fractions of the $f_1(1285)$ and $f_1(1420)$ produced in central pp interactions at 450 GeV/$c$," $Physics\ Letters\ B,$ vol. 440, no. 1-2, p. 225-232, 1998.
	\bibitem{ref25} D. Barberis et al. (WA102 Collaboration), "A study of the centrally produced $\pi^+\pi^-\pi^+\pi^-$ channel in pp interactions at 450 GeV/$c$," $Physics\ Letters\ B,$ vol. 413, no. 1-2, p. 217-224, 1997.
	\bibitem{ref26} D. Barberis et al. (WA102 Collaboration), "A spin analysis of the $4\pi$ channels produced in central pp interactions at 450 GeV/$c$," $Physics\ Letters\ B,$ vol. 471, no. 4, p. 440-448, 2000.
	\bibitem{ref27} D. Barberis et al. (WA102 Collaboration), "A study of the $KK\pi$ channel produced centrally in pp interactions at 450 GeV/$c$," $Physics\ Letters\ B,$ vol. 413, no. 1-2, p. 225-231, 1997.
	\bibitem{ref28} H. Y. Cheng and R. Shrock, "Some results on vector and tensor meson mixing in a generalized QCD-like theory," $Physical\ Review\ D,$ vol. 84, no. 9, p. 094008, 2011.
	\bibitem{ref29} D. M. Li, B. Ma, and H. Yu, "Regarding the axial-vector mesons," $The\ European\ Physical\  Journal\ A-Hadrons\ and\ Nuclei,$ vol. 26, no. 1, p. 141-145, 2005.
	\bibitem{ref30} F. E. Close and A. Kirk, "Implications of the Glueball-$q\bar{q}$ filter on the $1^{++}$ nonet," $Zeitschrift\ Physk\ C\ Particles\ and\ Fields,$ vol. 76, no. 3, p. 469-474, 1997.
	\bibitem{ref31} Ando, A., et al. "Experimental study of the axial-vector resonances of $a_1$ and $h_1$ in the $\pi - p$ charge exchange reaction," $Physics\ Letters\ B,$ vol. 291, no. 4, p. 496-502, 1992.
	\bibitem{ref32} X. C. Feng, F. C. Jiang, T. Q. Chang, and J. L. Feng, "Towards the understanding of 1$^1P_1$ meson mass spectrum," $Chinese\ Physics\ B,$ vol. 17, no. 12, p. 4472, 2008.
	\bibitem{ref33} K. B. Vijaya Kumar, Bhavyashri, Y. L. Ma, and A. Prakash, "P wave meson spectrum in a relativistic model with instanton induced interaction," arXiv preprint arXiv:0811.4308 (2008).
	\bibitem{ref34} S. Godfrey and N. Isgur, "Mesons in a relativized quark model with chromodynamics," $Physical\ Review\ D,$ vol. 32, no. 1, p. 189, 1985.
	\bibitem{ref35} P. V. Chliapnikov, "S-and P-wave meson spectroscopy in the nonrelativistic quark model," $Physics\ Letters\ B,$ vol. 496, no. 3-4, p. 129-136, 2000.
	\bibitem{ref36} J. Vijande, F. Fern¨¢ndez, and A. Valcarce, "Constituent quark model study of the meson spectra," $Journal\ of\ Physics\ G,$ vol. 31, no. 5, p. 481, 2005.
	\bibitem{ref37} M. Chizhov, "Vector meson couplings to vector and tensor currents in extended NJL quark model," $Journal\ of\ Experimental\ and\ Theoretical\ Physics\ Letters,$ vol. 80, no. 2, p. 73-77, 2004.
	\bibitem{ref38} T. Appelquist, J. Carazzone, "Infrared singularities and massive fields," $Physical\ Review\ D,$ vol. 11, no. 10, p. 2856, 1975.
	\bibitem{ref39} R. K. Carnegie et al, "$Q_1(1290)$ and $Q_2(1400)$ Decay Rates and their SU (3) Implications,"  $Physics\ Letters\ B,$ vol. 68, no. 3, p. 287-291, 1977.
	\bibitem{ref40} M. Suzuki, "Strange axial-vector mesons," $Physical\ Review\ D,$ vol. 47, no. 3, p. 1252, 1993.
	\bibitem{ref41} F. Divotgey, L. Olbrich and F. Giacosa, "Phenomenology of axial-vector and pseudovector mesons: decays and mixing in the kaonic sector," $The\ European\ Physical\ Journal\ A,$ vol. 49, no. 10, p. 135, 2013.
	\bibitem{ref42} H. Y. Cheng, "Revisiting axial-vector meson mixing," $Physics\ Letters\ B,$ vol. 707, no. 1, p. 116-120, 2012.
	\bibitem{ref43} H. G. Blundell, S. Godfrey and B. "Properties of the strange axial mesons in the relativized quark model," $Physical\ Review\ D,$ vol. 53, no. 7, p. 3712, 1996.
	\bibitem{ref44} L. Burakovsky and J. T. Goldman, "Constraint on axial-vector meson mixing angle from the nonrelativistic constituent quark model," $Physical\ Review\ D,$ vol. 56, no. 3, p. 1368, 1997.
	\bibitem{ref45} L. Burakovsky and J. T. Goldman, "Regarding the enigmas of P-wave meson spectroscopy," $Physical\ Review\ D,$ vol. 57, no. 5, p. 2879, 1998.
	\bibitem{ref46} A. Tayduganov, E. Kou and A. Le Yaouanc, "Strong decays of $K_1$ resonances," $Physical\ Review\ D,$ vol. 85, no. 7, p.074011, 2012.
	\bibitem{ref47} H. J. Lipkin, "A dynamical model for mixing of axial vector (Q) mesons," $Physics\ Letters\ B,$ vol. 72, no. 2, p. 249-250, 1977.
	
\end{thebibliography}
\end{document}